\documentclass[manuscript]{aastex}

\usepackage{hyperref}
\usepackage{url}
\usepackage{lscape}

\begin{document}

\title{A New Anisotropic Compact Star Model having Matese \& Whitman Mass Function}

\author{Piyali Bhar\altaffilmark{1}}
\affil{Department of Mathematics, Government General Degree College, Singur, Hooghly, West
Bengal-712409, India}
\email{piyalibhar90@gmail.com}

\author{B. S. Ratanpal\altaffilmark{2}}
\affil{Department of Applied Mathematics, Faculty of Technology \& Engineering, The M. S. University of Baroda, Vadodara - 390 001, India}
\email{bharatratanpal@gmail.com}

\begin{abstract}
A new singularity free model of anisotroipic compact star is proposed. 
The Einstein field equations are solved in closed form by utilizing Matese \& Whitman mass function. 
The model parameters $\rho$, $p_r$ and $p_t$ all are well behaved inside the stellar interior and our model satisfies 
all the required conditions to be physically acceptable. The model given in the present work is
compatible with observational data of compact objects like SAX J 1808.4-3658 (SS1), SAX J 1808.4-3658 (SS2) and 4U 1820-30.
A particular model of 4U 1820-30 is studied in detail and found that it satisfies all the condition needed for physically
acceptable model. The present work is the generalization of Sharma and Ratanpal (2013)  model for compact stars admitting quadratic
equation of state.
\end{abstract}


\section{Introduction}
The study of relativistic models of compact stars like neutron stars and strange stars
are always a topic of immense interest to the researchers in the field of astrophysics. Neutron
stars are composed of neutrons, while strange stars are
composed of quark or strange matter consisting of u, d,
and s quarks. \cite{Witten84} proposed that the formation of strange matter can
be classified into two ways: the quark hadron phase transition
in the early universe and conversion of neutron stars
into strange stars at ultrahigh densities. According to \cite{bodmar71} a phase transition between hadronic and 
strange quark matter may occur in the universe at a density higher than the nuclear density when a massive star explodes as a 
supernova. As a result strange quark is likely to be seen at the inner core of star. 
\cite{rud72} has shown that the fluid pressure of
the highly compact astrophysical objects like X-ray
pulsar, Her-X-1, X-ray buster 4U 1820-30, millisecond
pulsar SAXJ1804.4-3658 etc., whose density in central region
is expected to be beyond the nuclear density ($\sim 10^{15}
gm/cc)$ becomes anisotropy in nature, i.e. the pressure
inside the fluid sphere can be decomposed into two parts
radial pressure $p_r$ and transverse pressure $p_t$ . Their difference $\Delta=p_t-p_r$ is
called the measure of anisotropy.
Anisotropy may occurs in various reasons, e.g.,
the existence of solid core, in presence of type P superfluid,
phase transition, mixture of
two fluid, existence of external field etc.\\

A large number of anisotropic models are available in the literature. Local anisotropy in self-gravitating
systems were studied by \cite{herera97} (and reference within this). \cite{lobo06} has given a
model of a stable dark energy star by assuming two spatial
types of mass function: one is of constant energy density and
the other by considering \cite{mat80}  mass function. Various features of dark energy star
and stability analysis one can find in model proposed by \cite{lobo06}. Recently \cite{bhar15} proposed a 
new model of dark energy star
consisting of five zones, namely, the solid core of constant
energy density, the thin shell between core and interior, an
inhomogeneous interior region with anisotropic pressures, a
thin shell, and the exterior vacuum region. They discussed various
physical properties. The model satisfies all the physical
requirements and stability condition under a small radial
pulsations is also discussed. By assuming a special type of matter density \cite{dev04} proposed an model of 
anisotropic star. From their analysis authors have shown that the absolute stability bound $\frac{2M}{R}<\frac{8}{9}$ 
can be violated and the star's surface redshift may be arbitrarily large. 
A new class of interior solution of a (2+1)-dimensional anisotropic star in Finch and Skea spacetime corresponding to the 
exterior BTZ black hole was developed by \cite{bhar14a}. The model is obtained by considering the MIT bag model
Equation of state the metric potential $ g_{rr}$ proposed by \cite{fs89}. Singularity free quintessence star in
Krori-Barua spacetime are obtained by \cite{bhar14c}. Relativistic stellar model admitting a quadratic equation of state was 
proposed by \cite{sharma13}. The earlier work is generalized in modified Finch-Skea spacetime by \cite{panda14} by incorporating a dimensionless parameter n.\\\\
Motivated by the earlier work of Sharma \& Ratanpal \cite{sharma13}, in this present paper we have developd a new ansitropic model of 
strange star by using the \cite{mat80} mass function. The present work is the generalization of
work of \cite{sharma13}. The work is organized as follows: Sect. 2 contains, field equations and solution.
Exterior spacetime and matching conditions are discussed in Sect. 3. Sect. 4 contains, physical analysis of the model. 
Sect. 5 contains the conclusion.
\section{Field Equations and Solution}
We consider the static spherically symmetric spacetime metric for the interior of stellar configuration as
\begin{equation}\label{IMetric}
ds^{2}=-e^{\nu(r)}dt^{2}+e^{\lambda(r)}dr^{2}+r^{2}(d\theta^{2}+\sin^{2}\theta d\phi^{2}),
\end{equation}
and the energy-momentum tensor of the form
\begin{equation}\label{EMTensor}
T_{\nu}^{\mu}=(\rho+p_r)u^{\mu}u_{\nu}-p_t g_{\nu}^{\mu}+(p_r-p_t)\eta^{\mu}\eta_{\nu},
\end{equation}
with $ u^{i}u_{j} =-\eta^{i}\eta_j = 1 $ and $u^{i}\eta_j= 0$. Here the vector $u_i$ is the fluid 4-velocity and 
$\eta^{i}$ is the spacelike vector which is orthogonal to $ u^{i}$, $\rho$ is the matter density, $p_r$ and $p_t$ 
are respectively the radial and transverse pressure of the fluid.\\

The Einstein field equations corresponding to spacetime metric (\ref{IMetric}) and energy-momentum tensor (\ref{EMTensor})
with $G=1=c$ are given by
\begin{equation}\label{FE1}
e^{-\lambda}\left[\frac{\lambda'}{r}-\frac{1}{r^{2}} \right]+\frac{1}{r^{2}}=8\pi\rho,
\end{equation}
\begin{equation}\label{FE2}
e^{-\lambda}\left[\frac{1}{r^{2}}+\frac{\nu'}{r} \right]-\frac{1}{r^{2}}=8\pi p_r,
\end{equation}
\begin{equation}\label{FE3}
\frac{1}{2}e^{-\lambda}\left[ \frac{1}{2}\nu'^{2}+\nu''-\frac{1}{2}\lambda'\nu'+\frac{1}{r}(\nu'-\lambda')\right]=8\pi p_t.
\end{equation}

The mass function, m(r), within the radius `r' is given by,
\begin{equation}\label{mr}
m(r)=4\pi\int_0^{r}w^{2}\rho(w)dw.
\end{equation}
Using (\ref{mr}) Einstein field equations (\ref{FE1}) - (\ref{FE3}) becomes,
\begin{equation}\label{FE4}
e^{-\lambda}=1-\frac{2m}{r},
\end{equation}
\begin{equation}\label{FE5}
r(r-2m)\nu'=8\pi p_r r^{3}+2m,
\end{equation}
\begin{equation}\label{FE6}
\frac{4}{r}(8\pi \Delta)=8\pi(\rho+p_r)\nu'+2(8\pi p_r'),
\end{equation}
where $\Delta$ is measure of anisotropy. To solve the above set of equations (\ref{FE4}) - (\ref{FE6}) we take the mass function in a 
particular form
\begin{equation}\label{Mass}
m(r)=\frac{br^{3}}{2(1+ar^{2})},
\end{equation}
where `a' and `b' are positive constants. This particular type of mass function is known as \cite{mat80} 
mass function that gives a monotonic decreasing matter density, used earlier by \cite{mak03} to model an anisotropic
fluid star, \cite{lobo06} to develop a model of dark energy star, \cite{sharma07} to model a class of 
relativistic stars with a linear equation of state and \cite{maharaj08} to model a charged anisotropic 
matter with linear equation of state. Substituting (\ref{Mass}) into (\ref{FE4}) we get,
\begin{equation}\label{ELambda}
e^{\lambda}=\frac{1+ar^{2}}{1+(a-b)r^{2}},
\end{equation}
and the matter density can be obtained as,
\begin{equation}\label{rho1}
8\pi\rho=\frac{b(3+ar^{2})}{(1+ar^{2})^{2}}.
\end{equation}
Using the expression of m(r) given in equation (\ref{Mass}) into equation (\ref{FE5}) we obtain,
\begin{equation}\label{nudash}
\nu'=(8\pi p_r)\frac{(1+ar^{2})r}{1+(a-b)r^{2}}+\frac{br}{1+(a-b)r^{2}}.
\end{equation}

Now to integrate the equation (\ref{nudash}) we assume the radial pressure in the form
\begin{equation}\label{pr}
8\pi p_r=\frac{bp_0(1-ar^{2})}{(1+ar^{2})^{2}},
\end{equation}
where $p_{0}>0$. The expression of $p_r$ is reasonable due to the fact that $p_r$ is monotonic decreasing function of `r',
vanishes at $r=\frac{1}{\sqrt{a}}$ which gives the radius of the star. This choice of $p_r$ makes (\ref{nudash}) easily integrable.
The central pressure is giving by $\frac{bp_{0}}{8\pi}$.
Substituting (\ref{pr}) into (\ref{nudash}) we get,
\begin{equation}\label{nudash1}
	\nu'=\frac{\left(p_{0}+1 \right)br+\left(1-p_{0} \right)abr^{3}}{\left(1+ar^{2} \right)\left[1+\left(a-b \right)r^{2} \right]}.
\end{equation}

On integrating equation (\ref{nudash1}) we get,

\begin{equation}\label{nu}
	\nu=log\left\{C\left(1+ar^{2}\right)^{p_{0}}\left[1+\left(a-b \right)r^{2}\right]^{\left(b-2ap_{0}+bp_{0}\right)/\left(2a-2b\right)}  \right\},
\end{equation}
where C is the constant of integration,
and the spacetime metric in the interior is given by
\begin{equation}\label{IMetric1}
	ds^{2}=-C\left(1+ar^{2}\right)^{p_{0}}\left[1+\left(a-b \right)r^{2}\right]^{\left(b-2ap_{0}+bp_{0}\right)/\left(2a-2b\right)}dt^{2}+\left[\frac{1+ar^{2}}{1+\left(a-b \right)r^{2}}\right]dr^{2}+r^{2}\left(d\theta^{2}+\sin^{2}\theta d\phi^{2}\right).
\end{equation}

The anisotropic factor $\Delta=p_t-p_r$ is given by,
\footnotesize
\begin{equation}\label{delta}
	8\pi\Delta=\frac{r^{2}\left[\left(p_{0}^{2}b^{2}+4p_{0}b^{2}+3b^{2}-12abp_{0}\right)+\left(-2ab^{2}p_{0}^{2}+10ab^{2}p_{0}+4ab^{2}-8a^{2}bp_{0}\right)r^{2}+\left(a^{2}b^{2}+2a^{2}b^{2}p_{0}+p_{0}^{2}a^{2}b^{2}+4a^{3}bp_{0}\right)r^{4}\right]}{4\left(1+ar^{2}\right)^{3}\left[1+\left(a-b\right)r^{2}\right]}.
\end{equation}
\normalsize
The profile of $\Delta$ is shown in fig. 3. From the figure it is clear that the anisotropic factor vanishes at the center of
the star which is expected. 

The transverse pressure is then obtained as,
\begin{equation}\label{pt1}
	8\pi p_{t}=\frac{4bp_{0}+\left(p_{0}^{2}+3b^{2}-8abp_{0}\right)r^{2}+\left(-2ab^{2}p_{0}^{2}+10ab^{2}p_{0}+4ab^{2}-12a^{2}bp_{0}\right)^{4}+\left(a^{2}b^{2}+6a^{2}b^{2}p_{0}+p_{0}^{2}a^{2}b^{2}\right)r^{6}}{4\left(1+ar^{2}\right)^{3}\left[1+\left(a-b\right)r^{2}\right]}.
\end{equation}
From (\ref{rho1}), (\ref{pr}) and (\ref{pt1}), the expressions for $8\pi\frac{d\rho}{dr}$, $8\pi\frac{dp_{r}}{dr}$ and $8\pi\frac{dp_{t}}{dr}$ are respectively given by
\begin{equation}\label{drhodr}
	8\pi\frac{d\rho}{dr}=\frac{-2abr\left[5+ar^{2} \right]}{\left(1+ar^{2}\right)^{3}},
\end{equation}
\begin{equation}\label{dprdr}
	8\pi\frac{dp_{r}}{dr}=\frac{2abp_{0}r\left(ar^{2}-3 \right)}{\left(1+ar^{2} \right)^{3}},
\end{equation}
\begin{equation}\label{dptdr}
	8\pi\frac{dp_{t}}{dr}=\frac{r\left[A_{1}+A_{2}r^{2}+A_{3}r^{4}+A_{4}r^{6}+A_{5}r^{8} \right]}{2+B_{1}r^{2}+B_{2}r^{4}+B_{3}r^{6}+B_{4}r^{8}+B_{5}r^{10}+B_{6}r^{12}},
\end{equation}
where $A_{1}=b^{2}p_{0}^{2}+4b^{2}p_{0}-24abp_{0}+3b^{2}$, $A_{2}=-6ab^{2}p_{0}^{2}+36ab^{2}p_{0}-24a^{2}bp_{0}+2ab^{2}$,
$A_{3}=5ab^{3}p_{0}^{2}-10ab^{3}p_{0}+6a^{2}b^{2}p_{0}+24a^{3}bp_{0}+5ab^{3}-6a^{2}b^{2}$, 
$A_{4}=6a^{3}b^{2}p_{0}^{2}-6a^{2}b^{3}p_{0}^{2}+8a^{2}b^{3}p_{0}-32a^{3}b^{2}p_{0}+24a^{4}bp_{0}+6a^{2}b^{3}-6a^{3}b^{2}$,
$A_{5}=a^{3}b^{3}p_{0}^{2}-a^{4}b^{2}p_{0}^{2}+6a^{3}b^{3}p_{0}-6a^{4}b^{2}p_{0}+a^{3}b^{3}-a^{4}b^{2}$, $B_{1}=12a-4b$, 
$B_{2}=2b^{2}-20ab+30a^{2}$, $B_{3}=8ab^{2}-40a^{2}b+40a^{3}$, $B_{4}=12a^{2}b^{2}-40a^{3}b+30a^{4}$, $B_{5}=8a^{3}b^{2}-20a^{4}b+12a^{5}$ and
$B_{6}=2a^{4}b^{2}-4a^{5}b+2a^{6}$.

From (\ref{drhodr}) and (\ref{dprdr}) we get
\begin{equation}\label{dprdrho}
	\frac{dp_{r}}{d\rho}=\frac{p_{0}\left(3-ar^{2}\right)}{\left(5+ar^{2}\right)},
\end{equation}
and from (\ref{drhodr}) and (\ref{dptdr}) we get
\begin{equation}\label{dptdrho}
	\frac{dp_{t}}{d\rho}=\frac{\left(1+ar^{2}\right)^{3}\left[A_{1}+A_{2}r^{2}+A_{3}r^{6}+A_{4}r^{6}+A_{5}r^{8}\right]}{-ab\left(5+ar^{2}\right)\left[2+B_{1}r^{2}+B_{2}r^{4}+B_{3}r^{6}+B_{4}r^{8}+B_{5}r^{10}+B_{6}r^{12}\right]},
\end{equation}
where $A_{1},~A_{2},~A_{3},~A_{4},~A_{5},~B_{1},~B_{2},~B_{3},~B_{4},~B_{5}~ and~ B_{6}$ are described as above. 
\section{Exterior Spacetime and Matching Condition}
The interior spacetime metric (\ref{IMetric1}) should continously match with the exterior Schwarzschild spacetime metric at the
boundary $r=R$
(where $R$ is the radius of the star.) where $p_r(r=R)=0$. The exterior spacetime is given by the line element

\[ds^{2}=-\left(1-\frac{2M}{r}\right)dt^{2}+\left(1-\frac{2M}{r}\right)^{-1}dr^{2}\]
\begin{equation}
~~~~~~~~~~~~~~~~~~~~~+r^{2}\left(d\theta^{2}+\sin^{2}\theta d\phi^{2}\right),
\end{equation}
these matching conditions gives
\begin{equation}\label{a}
a=\frac{1}{R^{2}},
\end{equation}
\begin{equation}\label{b}
b=\frac{4M}{R^{3}},
\end{equation}
and constant of integration $C$ as
\begin{equation}\label{C}
	C=\left(1+aR^{2} \right)^{-\left(p_{0}+1 \right)}\left[1+\left(a-b \right)R^{2} \right]^{\left(2a-3b+2ap_{0}-bp_{0} \right)/\left(2a-2b\right)}.	
\end{equation}

The values of $a$ and $b$  are obtained for the strange stars SAXJ 1808.4-3658 (SS1) (Radius=7.07 km), SAXJ 1808.4-3658 (SS2) (Radius=6.35 km) and 4U~1820-30 (Radius=10 km) which are given in table 1.\\
\section{Physical Analysis}
For a physically acceptable model the solution must  satisfy the following conditions:\\

\begin{enumerate}
	\item[1.] The metric potential should be free from singularities inside the radius of the star moreover metric potential
                  should satisfy the conditions: $e^{\nu(r=0)}=$ some positive constant and $e^{-\lambda(r=0)}=1$ which are 
                  satisfied by our model. Spacetime metric (\ref{IMetric1}) indicates metric potentials are free from singularity in $0\leq r \leq R$. 
	\item[2.] The density $\rho$ and radial and tangential pressures $p_{r}$ and $p_{t}$ must be positive inside the stars.
		  From (\ref{rho1}) and (\ref{pr}), it is clear that density $\rho$ and radial pressure $p_{r}$ are positive throughout
                  the distribution. For tangential pressure to be positive throughout the distribution, following inequality must be
                  satisfied:
		  \begin{equation}\label{dptdrP}
			\left(1-b^{2}\right)p_{0}^{2}+\left(-16ab+16b^{2}\right)p_{0}+8b^{2}\geq 0.
                  \end{equation}
	\item[3.] The radial pressure $p_r$ must be vanishing but the tangential pressure $p_t$  may not necessarily vanish at 
		  the boundary $r=R$ of the fluid sphere. However, the radial pressure is equal to the tangential pressure at the 
		  center of the fluid sphere i.e., $\Delta=0$ at the origin. Our model satisfies these conditions.
	\item[4.] Density, radial pressure and tangential pressure must be decreasing in radially outward direction. From (\ref{drhodr})
		  and (\ref{dprdr}), it is clear that $\frac{d\rho}{dr}\leq 0$ and $\frac{dp_{r}}{dr}\leq 0$ throughout the distribution.
		  From (\ref{dptdr}), $\frac{dp_{t}}{dr}\leq 0$ if 
		  \begin{equation}\label{p01}
			p_{0}<\frac{11a-12b}{8a+4b}.
		  \end{equation}
	\item[5.] For a physically acceptable model $\frac{dp_{r}}{d\rho}$ and $\frac{dp_{t}}{d\rho}$ must be less than 1.
		  $\frac{dp_{r}}{d\rho}<1$ at the centre restricts $p_{0}$ by
		  \begin{equation}\label{dprdrho0}
			p_{0}<\frac{5}{3},
		  \end{equation} 
		  and $\frac{dp_{r}}{d\rho}<1$ at the surface of the star restricts $p_{0}$ by
		  \begin{equation}\label{dprdrhoR}
			p_{0}<3.
		  \end{equation}
		  The condition (\ref{dprdrho0}) is much stronger than (\ref{dprdrhoR}). $\frac{dp_{t}}{d\rho}<1$ at the centre 
		  leads to inequality
		  \begin{equation}\label{dptdrho0}
			b^{2}p_{0}^2+\left(4b^{2}-24ab\right)p_{0}+3b^{2}+10ab<0,
		  \end{equation}
		  and $\frac{dp_{t}}{d\rho}<1$ at the boundary of stellar configuration gives inequality
		  \begin{equation}\label{dptdrhoR}
			-6ab^{2}p_{0}^{2}+\left(14ab^{2}+4b^{3}\right)p_{0}+12b^{3}+69a^{2}b+88ab^{2}-24b^{3}<0.
		  \end{equation}
	\item[6.] For a fluid sphere the trace of the energy tensor should be positive (Bondi\cite{bondi99}), i.e.
		  \begin{equation}\label{trace}
			\rho-p_{r}-2p_{t}\geq 0.
		  \end{equation}
	\item[7.] For an anisotropic fluid sphere all the energy conditions, namely Weak Energy Condition (WEC), Null Energy Condition (NEC), Strong Energy Condition (SEC) and Dominant Energy Condition (DEC) are satisfied if and only if the following inequalities hold simultaneously in every point inside the fluid sphere.
		  \begin{equation}\label{NEC}
(i)NEC:\rho+p_r\geq 0,
\end{equation}
\begin{equation}\label{WEC}
(ii)WEC:p_r+\rho\geq 0,~~~\rho>0,
\end{equation}
\begin{equation}\label{SEC}
(iii)SEC:\rho+p_r\geq 0~~~~\rho+p_r+2p_t\geq 0,
\end{equation}
\begin{equation}\label{DEC}
(iv)DEC:\rho >\left|p_r\right| ~~~, \rho >\left|p_t\right|.
\end{equation}
	\item[8.] The equation of state parameter $\omega_r$ is given by
		  \begin{equation}\label{EOSPara}
			\omega_r=\frac{p_r}{\rho}.
		  \end{equation}

\end{enumerate}

\begin{figure}[htbp]
    \centering
        \includegraphics[scale=.3]{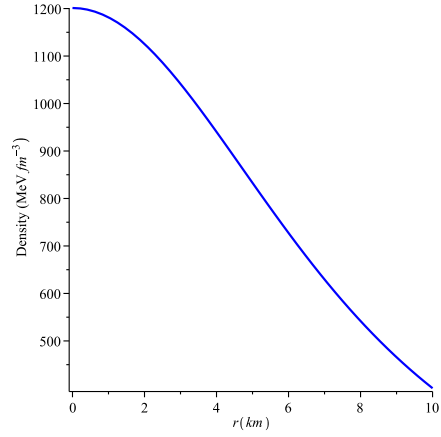}
       \caption{The matter density is plotted against `r' in the interior of the strange star 4U 1820-30}
    \label{fig:1}
\end{figure}

\begin{figure}[htbp]
    \centering
        \includegraphics[scale=.3]{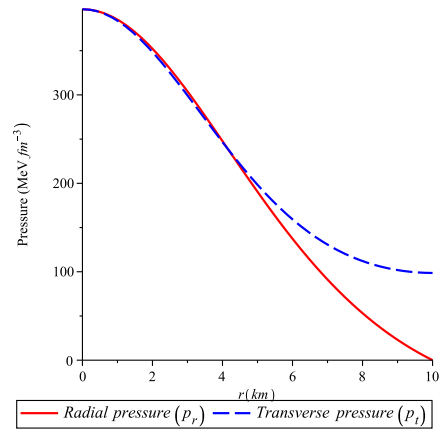}
       \caption{The radial and transverse pressure are plotted against `r' for the strange star 4U 1820-30}
    \label{fig:2}
\end{figure}

\begin{figure}[htbp]
    \centering
        \includegraphics[scale=.3]{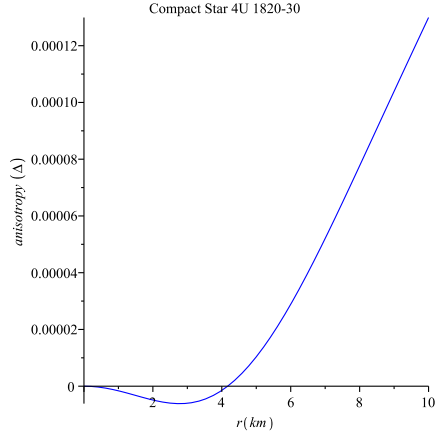}
       \caption{The anisotropy factor $\Delta=p_t-p_r$  in the interior of the strange star 4U 1820-30 is shown against `r'}
    \label{fig:3}
\end{figure}

\begin{figure}[htbp]
    \centering
        \includegraphics[scale=.3]{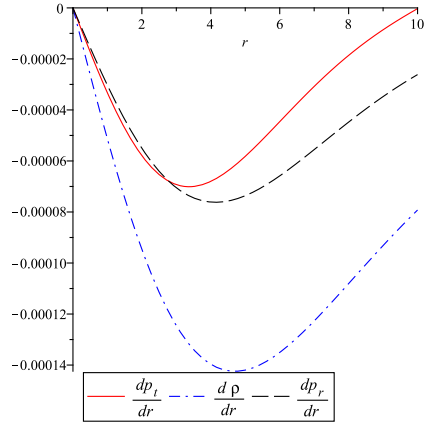}
       \caption{$\frac{d\rho}{dr}$,$\frac{dp_r}{dr}$ and $\frac{dp_{t}}{dr}$ are plotted against `r' for the strange star 4U 1820-30}
    \label{fig:4}
\end{figure}

\begin{figure}[htbp]
    \centering
        \includegraphics[scale=.3]{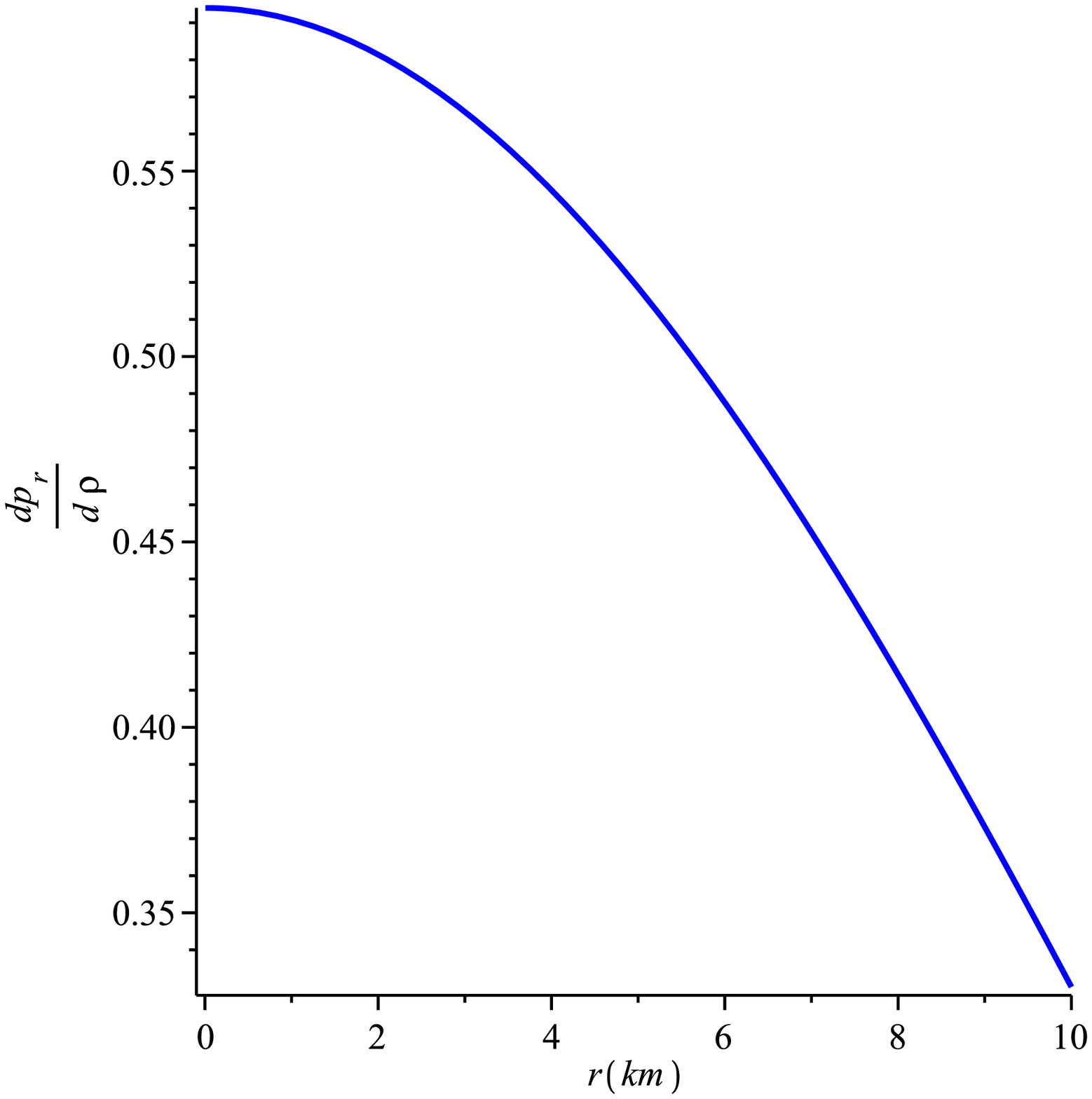}
       \caption{$\frac{dp_r}{d\rho}$ is plotted against `r' for the strange star 4U 1820-30.}
    \label{fig:5}
\end{figure}

\begin{figure}[htbp]
    \centering
        \includegraphics[scale=.3]{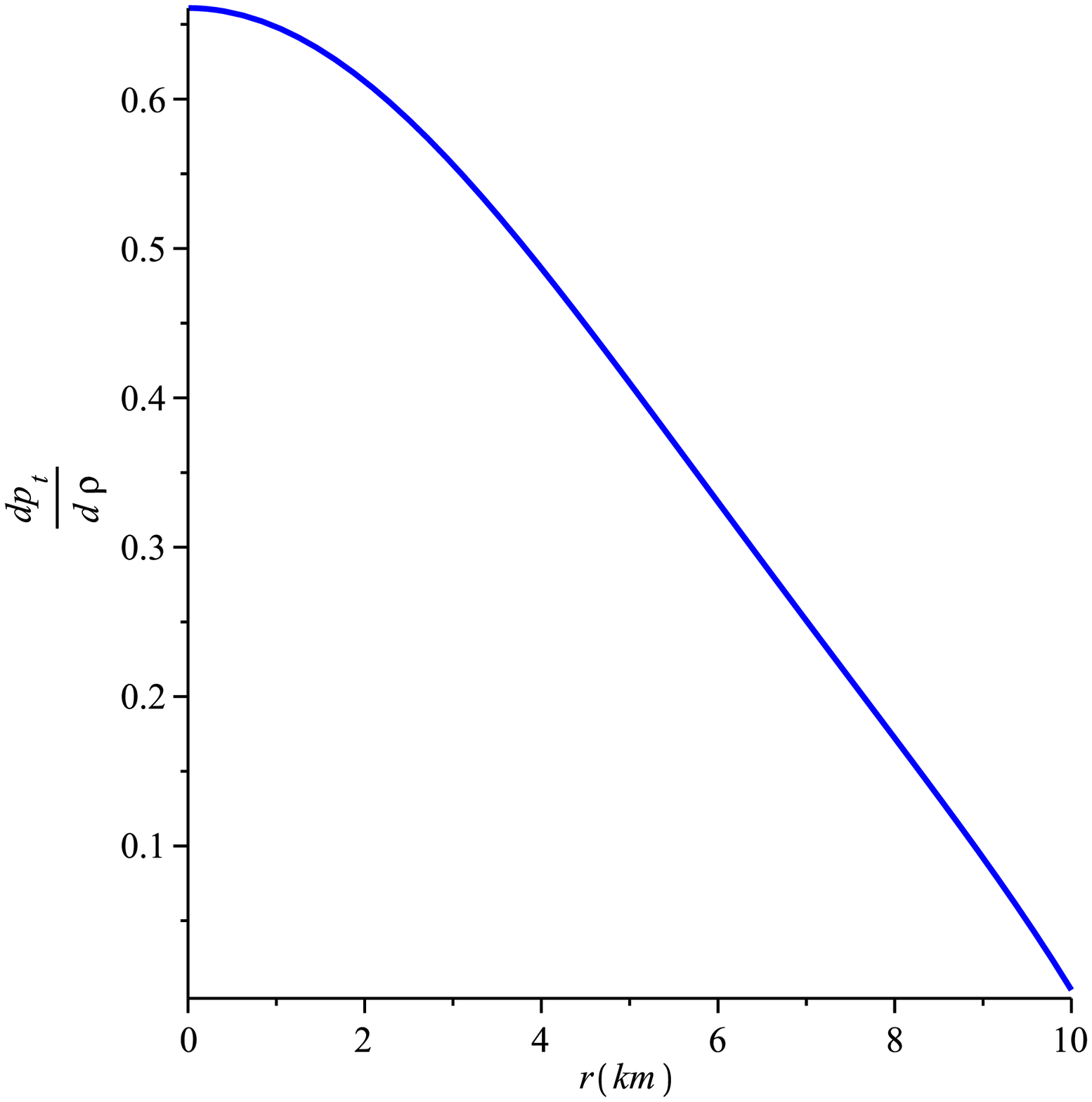}
       \caption{$\frac{dp_t}{d\rho}$ is plotted against `r' for the strange star 4U 1820-30.}
    \label{fig:6}
\end{figure}

\begin{figure}[htbp]
    \centering
        \includegraphics[scale=.3]{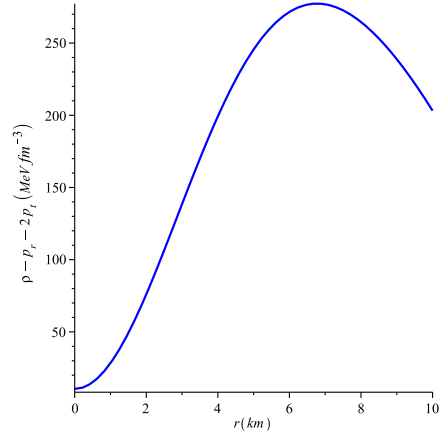}
       \caption{ $\rho-p_r-2p_t$ is shown against `r' for the strange star 4U 1820-30}
    \label{fig:7}
\end{figure}

\begin{figure}[htbp]
    \centering
        \includegraphics[scale=.3]{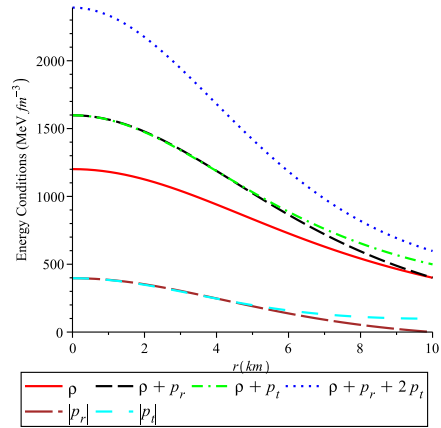}
       \caption{The energy conditions are plotted against `r' for the strange star 4U 1820-30.}
    \label{fig:8}
\end{figure}

\begin{figure}[htbp]
    \centering
        \includegraphics[scale=.3]{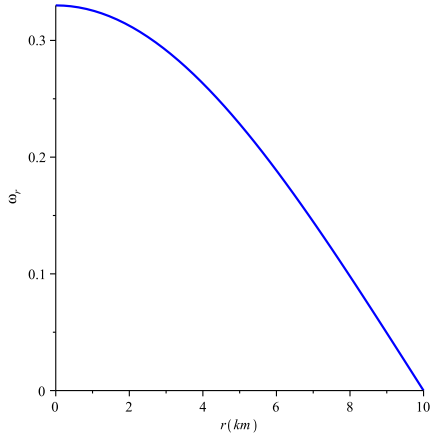}
       \caption{Equation of state parameters $\omega_r$ against `r' for the strange star 4U 1820-30. The plot indicates that $0\leq \omega_r\leq 1 $}
    \label{fig:9}
\end{figure}

For a physically acceptable model we need to choose values of $a$, $b$ and $p_{0}$ such that conditions (\ref{dptdrP}),
(\ref{p01}), (\ref{dprdrho0}), (\ref{dptdrho0}), (\ref{dptdrhoR}), (\ref{trace}), (\ref{NEC}), (\ref{WEC}), (\ref{SEC}) and
(\ref{DEC}) must be satisfied. For $a=0.01\;km^{-2}$, $b=0.013275\;km^{-2}$ and $p_{0}=0.99$, we have done graphical analysis
for the star $4U 1820-30$. It can be seen that all the above mentioned conditions are satisfied. The profiles of $\rho$, $p_{r}$ and $p_{t}$ are shown in fig. 1 and fig. 2 respectively.
From these figures it is clear that $\rho$, $p_{r}$ and $p_{t}$ are positive inside the fluid sphere. 
The graph of anisotropic parameter $\Delta$ is shown in fig. 3. Fig. 4 shows that $\rho$,
$p_{r}$ and $p_{t}$ are decreasing in radially outward direction. It can be seen that $\frac{dp_{r}}{d\rho}<1$ and $\frac{dp_{t}}{d\rho}<1$
from fig. 5 and fig. 6. Fig. 7 shows $\rho-p_{r}-2p_{t}\geq 0$. From fig. 8 it is clear that weak energy condition, null energy condition,
strong energy condition and dominant energy condition are satisfied throught the distribution. It can be seen from fig. 9 that
$0\leq \omega_{r}\leq 1$, hence the underlying matter is non-exotic in nature.\\\\
The surface redshift $z_s(r_b)$ of a star is given by the formula\\
\[1+z_s(r_b)=(1-2u(r_b))^{-\frac{1}{2}}.\]
Therefore $z_s$ can be obtained as,
\begin{equation}\label{zs}
z_s(r_b)=\left(\frac{1+(a-b)r_b^{2}}{1+ar_b^{2}}\right)^{-\frac{1}{2}}-1.
\end{equation}

Plugging the values of `a' and `b' the value of the surface redshift for strange stars SAXJ1808.4-3658 (SS1) (Radius=7.07 km.), 
SAXJ1808.4-3658 (SS2) (Radius=6.35 km) and 4U~1820-30 (Radius=10 km) are obtained from the equation (\ref{zs}) which is given in 
table-2.  In this respect we want to mentioned that according to \cite{mak06} for an anisotropic star in the presence of a 
cosmological constant the surface redshift should lie in the range $z_s \leq 5$ . From the table-2 it is clear that the maximum 
value of the surface redshift lies in the range proposed by \cite{mak06}.\\
\begin{table}[h]
\caption{The values of $a$ and $b$ obtained from the equation (\ref{a}) and (\ref{b})}
\label{tab:1}
\begin{tabular}{|l|l|l|l|l|l|}
\hline\noalign{\smallskip}
\textbf{Compact Star} & \textbf{$ M\;(M_\odot) $} & \textbf{ Mass (km) } & \textbf{Radius (km)} & \textbf{$a\;(km^{-2})$} & \textbf{$ b\;(km^{-2}) $} \\
\noalign{\smallskip}\hline\noalign{\smallskip}
\textbf{SAX J 1808.4-3658(SS1)} 	  & 1.435 & 2.116625  & 7.07   & 0.02 & 0.02395773027 \\
\textbf{SAX J 1808.4-3658(SS2)} 	  & 1.323 & 1.951425 & 6.35 & 0.0248 & 0.03048531450  \\
\textbf{4U 1820-30} 	  & 2.25 & 3.31875  &  10 & 0.01 & 0.01327500000  \\
\noalign{\smallskip}\hline
\end{tabular} 
\end{table}

\begin{table}[h]
\caption{The values of central density, surface density, maximum value of $\frac{2M}{r_b}$ and maximum surface redshift}
\label{tab:2}
\begin{tabular}{|l|l|l|l|l|}
\hline\noalign{\smallskip}
\tiny\textbf{Compact Star} & \tiny\textbf{Central density ($\rho_{0}$) gm/cc} & \tiny\textbf{ Surface density ($\rho_{s}$) gm/cc } & \tiny\textbf{$\frac{2M}{b}$ Max. Value} & \tiny\textbf{$z_{s}$ Max. Value}\\
\noalign{\smallskip}\hline\noalign{\smallskip}
\tiny\textbf{SAX J 1808.4-3658(SS1)} 	  & \tiny$3.860681427 \times 10^{15}$ &  \tiny$1.286893809 \times 10^{15}$ &  \tiny0.5987623760$<\frac{8}{9}$  & \tiny 0.578698422 \\
\tiny\textbf{SAX J 1808.4-3658(SS2)} 	  & \tiny$4.912572524 \times 10^{15}$ & \tiny$1.637524175 \times 10^{15}$ &\tiny 0.6146220470$<\frac{8}{9}$ & \tiny 0.610855437  \\
\tiny\textbf{4U 1820-30} 	  & \tiny$2.139207068 \times 10^{15}$ & \tiny$0.713069023 \times 10^{15}$  & \tiny0.6637500000 $<\frac{8}{9}$  &  \tiny0.724522454  \\
\noalign{\smallskip}\hline
\end{tabular} 
\end{table}

\section{Conclusion}

Utilizing \cite{mat80} mass function we have obtained a new analytical model of compact star corresponding 
to the exterior Schwarzschild spacetime. The model parameters `a' and `b' are obtained from some physical conditions. The matter density, 
radial and transverse pressure all are well behaved inside 
the stellar configuration and radial pressure vanishes at the boundary of the star. The anisotropy vanishes at the centre of the star.
The salient features of present model is: If we take $a=b=\frac{1}{R^{2}}$, where $R$ is curvature parameter, 
our result coincides with the result of \cite{sharma13}. Hence our model is generalization of model of superdense
star proposed by \cite{sharma13}.

\section*{Acknowledgement}
BSR would like to thank IUCAA, Pune for providing the facilities and hospitality for carrying out this work.


\end{document}